# Absence of signatures of Weyl orbits in the thickness dependence of quantum transport in cadmium arsenide

Luca Galletti[1,*], Timo Schumann[1], David A. Kealhofer[2], Manik Goyal[1], and Susanne Stemmer[1]

[1]Materials Department, University of California, Santa Barbara, CA 93106-5050, USA.

[2]Department of Physics, University of California, Santa Barbara, CA 93106-9530, USA.

[*] Email: luca_galletti@ucsb.edu

**Abstract**

In a Weyl orbit, the Fermi arc surface states on opposite surfaces of the topological semimetal are connected through the bulk Weyl or Dirac nodes. Having a real-space component, these orbits accumulate a sample-size-dependent phase. Following recent work on the three-dimensional Dirac semimetal cadmium arsenide ($Cd_3As_2$), we have sought evidence for this thickness-dependent effect in quantum oscillations and quantum Hall plateaus in (112)-oriented $Cd_3As_2$ thin films grown by molecular beam epitaxy. We compare quantum transport in films of varying thickness at apparently identical gate-tuned carrier concentrations and find no clear dependence of the relative phase of the quantum oscillations on the sample thickness. We show that small variations in carrier densities, difficult to detect in low-field Hall measurements, lead to shifts in quantum oscillations that are commensurate with previously reported phase shifts. Future claims of Weyl orbits based on the thickness dependence of quantum transport data require additional studies that demonstrate that these competing effects have been disentangled.

A Weyl orbit is a unique transport phenomenon of topological semimetals. It involves the Fermi arcs that exist on opposite surfaces of a Weyl semimetal and a connection though the bulk nodes [1]. The resulting closed orbit can give rise to quantum oscillations in sufficiently strong magnetic fields. Three-dimensional Dirac semimetals, such as $Cd_3As_2$ [2], are also thought to host Weyl orbits, because each Dirac node can be considered as two coincident Weyl nodes of opposite chirality [1].

While angle-resolved x-ray photoemission has provided clear evidence of Fermi arcs [3-5], detecting signatures of Weyl orbits has proven to be more challenging. Experimental reports in support of Weyl orbits include two-dimensional Shubnikov-de Haas oscillations of thin slabs of $Cd_3As_2$ and their disappearance when the samples were shaped into triangles [6]. The triangular shape is expected to lead to destructive interference of individual Weyl orbits. Recently, the quantum Hall effect has been observed in thin films and platelets of $Cd_3As_2$ [7-10], but the nature of the electronic states that give rise to it has not yet been fully clarified. Several different interpretations have been put forward in the literature, including quantum confined bulk states [2,8], topological surface states [11], and Weyl orbits [10]. To support the latter interpretation, a recent study has investigated the thickness dependence of the quantum Hall effect of wedge-shaped platelets of $Cd_3As_2$ [12]. In the thinner part of the wedge, the quantum Hall plateaus appeared at a slightly lower magnetic field than in a thicker region. This observation appears consistent with expectations of a thickness-dependent term in the modified Lifshitz-Onsager relation for quantum oscillations involving Weyl orbits, where conductivity minima appear at magnetic fields $B$ given by [1,12]:

$$\frac{1}{B} = \frac{2\pi e}{\hbar c A_N} \left( n + \gamma - \frac{L}{2\pi}(k_w + 2k_F) \right), \quad (1)$$

where $e$ is the free electron charge, $\hbar$ is the reduced Planck's constant, $c$ is the speed of light, $A_N$ is the area enclosed by electrons in $k$-space with their cyclotron orbits on the Fermi surface, $n$ is the Landau level index, $\gamma$ is the Berry phase, $L$ is the sample thickness, $k_w$ is the separation of the Weyl/Dirac nodes along the direction of $B$, and $k_F$ is the Fermi wave vector. Here, the additional phase term, $L(k_w + 2k_F)/2\pi$, is related to the tunneling of electrons from the top to the bottom surface [1]. The experimental signature of this additional phase is its dependence on the thickness $L$.

As discussed in ref. [1], there are many reasons why Weyl orbits may not be easily observed in Dirac semimetals, such as $Cd_3As_2$. For example, the bulk nodes are expected to be gapped by quantum confinement in thin slabs for which the quantum Hall effect has been reported [7,9,11,13]. Depending on the surface potential [14], Fermi arcs may morph into topological surfaces states that not necessarily remain connected to the bulk nodes [1,15]. In the conventional (112) surface orientation, with $B$ normal to the surface, the four-fold symmetry that protects the bulk nodes is broken. We note that the presence of topological surface states is required, however, by the $\mathbb{Z}_2$ invariant, even when the bulk nodes are gapped [2].

The goal of the present paper is not to address these issues, but to look more closely at the sensitivity of the quantum oscillations of $Cd_3As_2$ to the sample thickness and carrier density. In particular, the frequency of Shubnikov-de Haas oscillations is very sensitive to small variations in carrier density, which, in real samples, may also depend on the sample thickness. To this end, we use a gate voltage, which is applied to a gated Hall bar structure, to match the carrier density to that of four additional Hall bars made from epitaxial $Cd_3As_2$ thin films having different thicknesses. This allows us to disentangle the effects of charge carrier density from a true thickness dependence of the quantum oscillations, as may arise from Weyl orbits. We show that the phase of the Shubnikov-de Haas oscillations measured from thin films of $Cd_3As_2$ lacks a clear correlation with the film thickness and that it is reasonable to expect the same to be true for other kinds of realistic $Cd_3As_2$ specimens.

Epitaxial $Cd_3As_2$ thin films were grown by molecular beam epitaxy on (111)GaSb/GaAs substrates, as described in detail elsewhere [16]. Film thicknesses were determined via x-ray reflectivity performed on a Panalytical MRD PRO Materials Research Diffractometer, using Cu K-alpha radiation and an analyzer crystal. The data were fit to a one-layer model using the GenX software package [17]. Thickness maps recorded for films grown on 2-inch wafers show that the thickness in a 1×1 $cm^2$ area (i.e. the size of the samples in the presented study) varies by about 0.7 nm, a variance larger than any error obtained from the fitting procedure. These films show a quantum Hall effect within a relatively wide thickness range (10 – 60 nm) [9] that originates from two-dimensional Dirac fermions [11], which we have previously interpreted to likely reside in topological surface states. A gap in the bulk spectrum is expected to open for films below a certain thickness and, consistent with this, parallel conduction from bulk-like carriers vanishes in films thinner than 60 nm at sufficiently low temperatures. Here, we compare the magnetotransport of (112)-oriented $Cd_3As_2$ films of different thicknesses. In addition, a gated 100 μm Hall bar structure was fabricated from a 27-nm-thick $Cd_3As_2$ film, using a 37 nm of $Si_2ON_2$ gate dielectric deposited by ion beam deposition after performing an in-situ $N^*$ plasma treatment [14]. Magnetotransport measurements were carried out in a Quantum Design Physical Properties Measurement System (PPMS) at 2 K and magnetic fields up to 9 T. The longitudinal ($R_{xx}$) and Hall ($R_{xy}$) magnetoresistances were measured using a current bias of 1 μA.

Figure 1 shows the results from the magnetoresistance measurements. In the left panels, the background-subtracted $R_{xx}$ of four Hall bars, each from a film with different thickness, is shown as a function of *B*, together with results from the gated Hall bar on the 27-nm film. The corresponding Hall data is shown in the right panels. In each panel, the gate bias of the gated Hall bar is chosen so that its carrier density matches that of the other sample shown. To match their carrier densities, the criterion that is employed here is the alignment of the low-field $R_{xy}(B)$ curves, so that they have the same slope (c.f. right panels of Fig. 1). We note that this is same approach that was used in ref. [12].

All samples show pronounced Shubnikov-de Haas oscillations above 4 T, demonstrating comparable quantum mobilities, which are at least as high as those in ref. [12]. The twofold degeneracy of the Landau levels is lifted in magnetic fields close to 7 T [7]. For this reason, we focus our discussion on oscillations below 7 T. As indicated by the dashed lines in the left panels, in some cases the oscillations are perfectly aligned despite a substantial difference in the film thickness [Figs. 1(a) and (d)]. In Fig. 1(c) the thinner film approaches the quantum Hall plateau at lower fields, as in ref. [12]. In Fig. 1(b) there is a substantial discrepancy, despite near identical film thicknesses. In short, the measurements lack a clear correlation of the phase of the Shubnikov–de Haas oscillations with film thickness.

We attribute this to the fact that even small differences in carrier density, too small to be detected in the low field Hall data, produce a sizable shift in high-field Shubnikov-de Haas oscillations and quantum Hall plateaus. In other words, aligning the low field Hall data is insufficient to ensure that the carrier densities are identical, given the size of the error involved. To illustrate this problem, we show in Fig. 2(a) simulations using the standard Lifshitz-Onsager relation at 300 mK for a material with an effective mass $0.02m_e$ [9,18], where $m_e$ is the bare electron mass, and an elastic scattering time of 10 fs. We note that these parameters only affect the damping of the amplitude of the oscillations and are thus not critical for the purpose of this discussion. Shown are simulations for two slightly differing carrier densities, $0.95\times10^{12}$ cm$^{-2}$ and $1.00\times10^{12}$ cm$^{-2}$, i.e., typical two-dimensional carrier densities of thin slabs of Cd$_3$As$_2$ [7]. At high fields, the shift between the two sets of oscillations increases. While the determination of the carrier density from the Hall coefficient, i.e., $R_{xy}(B)$, shown in Fig. 2(b), should, in principle, not be limited to low fields, the onset of the quantum Hall effect limits the field range for estimations of the carrier density in practice. Furthermore, Dirac systems often contain $p$-type carriers, which produce a curvature in the high-field $R_{xy}(B)$ data, see, e.g., ref. [14]. An example is Fig. 1(e), where it is responsible for the discrepancy in the low-field Hall data. In the example shown in Fig. 2, a 5% variation in carrier density produces a shift in the oscillations of 2 T at a field of 30 T, but only a difference in the value of $R_{xy}$ of 0.0064 $h/e^2$ at 5 T, a change so small that it would be completely undetectable in the data in Fig. 1 or ref. [12]. The shift between the oscillation maxima in the data Fig. 1(c) is about 300 mT. Rather than invoking Weyl orbits, this shift can be explained by a variation of the carrier density of about 4% between the two samples, which should produce a change $\Delta R_{xy} = 10$ $\Omega$ at 0.4 T and 50 $\Omega$ at 2 T (at the onset of quantum oscillations) in the Hall data, which falls within the uncertainties of the low field Hall data. To illustrate this point, the small dot in Fig. 1(g) corresponds to this $\Delta R_{xy}$. Thus, small variations in the carrier densities between samples of different thickness cannot be disregarded as a source of an apparently thickness-dependent phase shift in the Shubnikov-de Haas oscillations.

In conclusion, based on the available data, the high degree of sensitivity of the positions of the maxima in the Shubnikov-de Haas oscillations and their associated Hall plateaus to small variations in the carrier density, in combination with the relative insensitivity of the low-field Hall data to such variations, does not allow us to invoke Weyl orbits based on thickness data for realistic samples. There are many reasons why carrier densities may not be completely independent of the sample thickness. For example, the growth rate, and therefore the incorporation of defects that cause the unintentional doping, may differ. Incidentally, we note that a spatial variation in growth rate would give rise to a wedge-shaped sample. Another reason is the sensitivity of the $Cd_3As_2$ surface potential to the physical condition of the (air-exposed) surface [14]. The resulting band bending effects will produce a gradient in the carrier density. Finally, we wish to emphasize that the analysis presented here does not disprove the existence of Weyl orbits in general.

## Acknowledgements

The authors thank Tess Winkelhorst for help with acquiring some of the data shown. They also gratefully acknowledge support through the Vannevar Bush Faculty Fellowship program by the U.S. Department of Defense (grant no. N00014-16-1-2814). This work made use of the MRL Shared Experimental Facilities, which are supported by the MRSEC Program of the U.S. National Science Foundation under Award No. DMR 1720256.

**Figure Captions**

**Figure 1:** Magnetoresistance data from $Cd_3As_2$ thin films with different thickness. Also shown in each panel is data from a gated device with a 27-nm-thick $Cd_3As_2$ film at different values of the applied gate voltage (blue curve). The labels in the graphs indicate the film thickness and the gate voltage, respectively. (a-d) Longitudinal resistance $R_{xx}$ as a function of magnetic field ($B$), after background subtraction, showing Shubnikhov-de Haas oscillations. The lines indicate the relative alignments of prominent maxima between the two sets of data in each panel. (e-h) Transverse resistance ($R_{xy}$) showing quantum Hall plateaus. The carrier density of the gated device is matched to that of the other film shown via the gate voltage, using identical slopes of $R_{xy}(B)$ as the matching criterion. The inset shows a magnification of the data at the lowest fields (0.4 T). In (g) and its inset, we show a small dot (see arrows) that depicts a 10 Ω and 50 Ω variation of the $\Delta R_{xy}$ at 0.4 and 2 T, respectively, as would be caused by variation of the carrier density of 4% (see text for details).

**Figure 2:** (a) Simulated Shubnikov-de Haas oscillations for two different carrier densities. A 5% difference in the carrier density produces a sizable shift of the position of the Shubnikov-de Haas oscillation at high fields (around 2 T, as depicted by the dashed lines). The same difference produces only a small change in the low-field Hall resistance ($R_{xy}$), shown in (b), as depicted by the dashed lines.

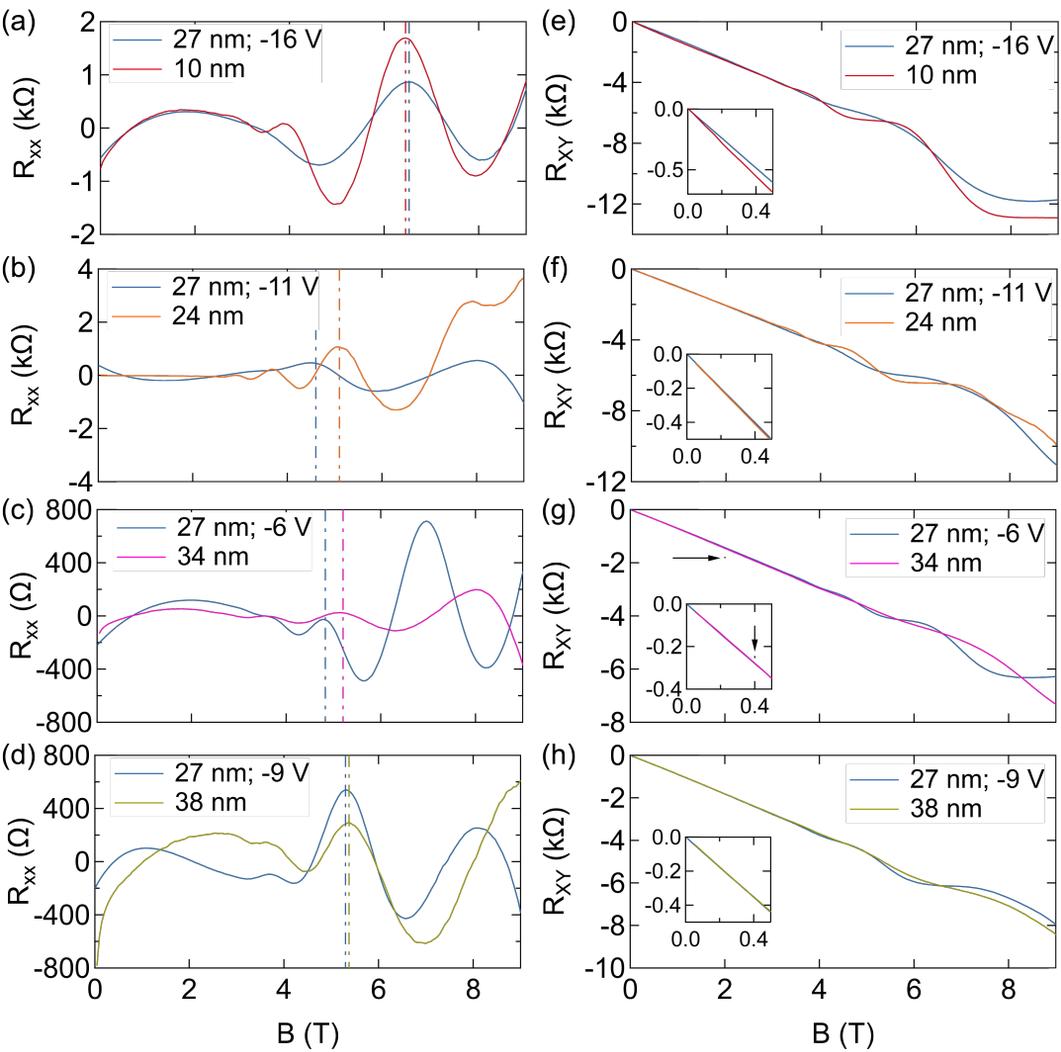

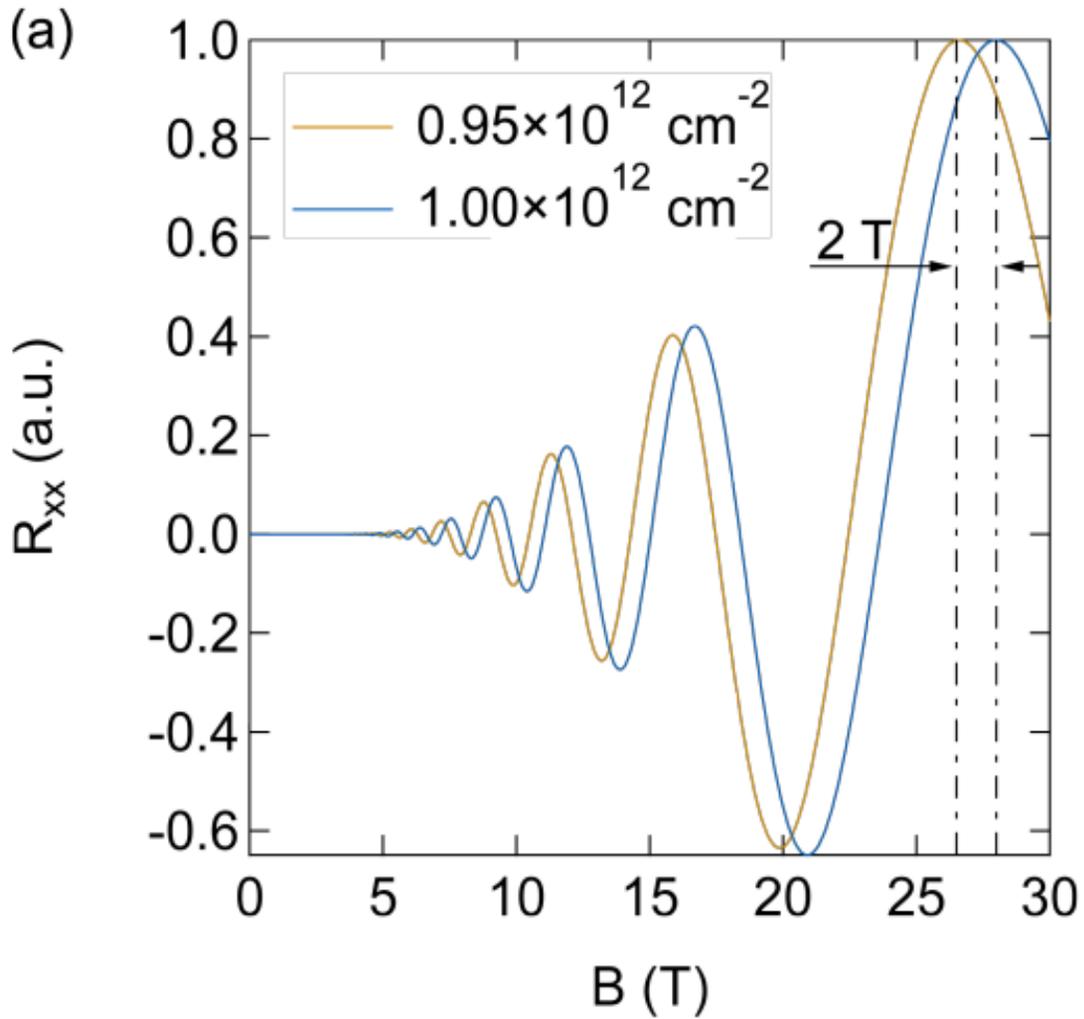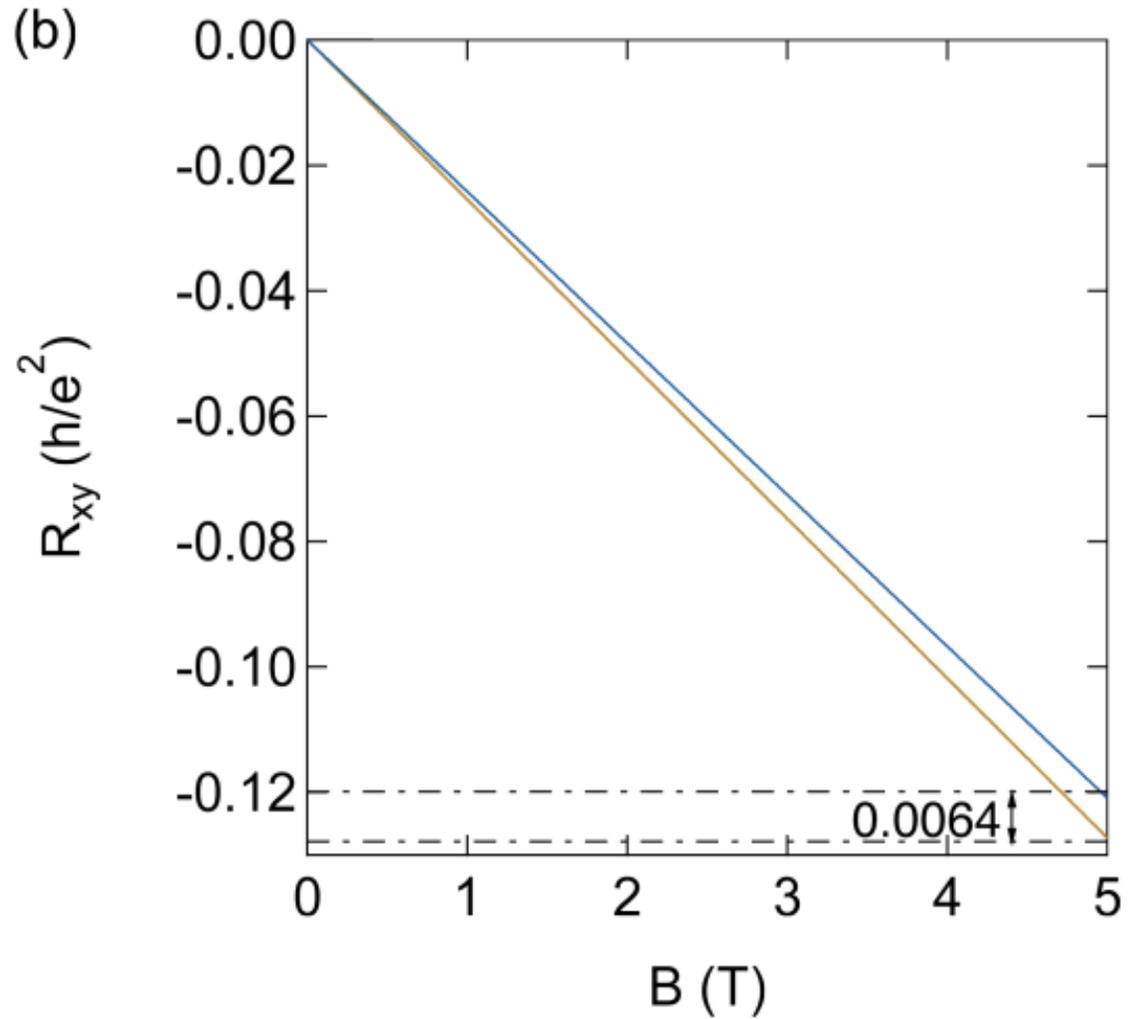